\documentclass{article}

\PassOptionsToPackage{numbers, compress}{natbib}
\usepackage[preprint]{neurips2024}

\usepackage[utf8]{inputenc} 
\usepackage[T1]{fontenc}    
\usepackage{hyperref}       
\usepackage{url}            
\usepackage{booktabs}       
\usepackage{amsfonts}       
\usepackage{nicefrac}       
\usepackage{microtype}      
\usepackage{xcolor}         

\title{Troubling Taxonomies in GenAI Evaluation}

\author{%
  Glen Berman\\
  Australian National University\\
  Canberra, ACT, Australia\\
  \texttt{glen.berman@anu.edu.au}\\
  \And
  Ned Cooper\\
  Australian National University\\
  Canberra, ACT, Australia\\
  \texttt{edward.cooper@anu.edu.au}\\
  \And
  Wesley Hanwen Deng\\
  Carnegie Mellon University\\
  Pittsburgh, Pennsylvania, USA\\
  \texttt{hanwend@andrew.cmu.edu}\\
  \And
  Ben Hutchinson\\
  Google Research\\
  Pyrmont, NSW, Australia\\
  \texttt{benhutch@google.com}\\
}

\begin{document}

\maketitle


\begin{abstract}
    To evaluate the societal impacts of GenAI requires a model of how social harms emerge from interactions between GenAI, people, and societal structures. Yet a model is rarely explicitly defined in societal impact evaluations, or in the taxonomies of societal impacts that support them. In this provocation, we argue that societal impacts should be conceptualised as application- and context-specific, incommensurable, and shaped by questions of social power. Doing so leads us to conclude that societal impact evaluations using existing taxonomies are inherently limited, in terms of their potential to reveal how GenAI systems may interact with people when introduced into specific social contexts. We therefore propose a governance-first approach to managing societal harms attended by GenAI technologies.
\end{abstract}

\section{Models of the social world in GenAI evaluations}

Recent scholarship has started to make explicit the normative values and commitments of GenAI and Machine Learning (ML) practices \cite{birhane2022values} and evaluation \cite{hutchinson2022evaluation}. 
In this provocation, we extend this line of inquiry by arguing that we need to attend to the implicit values and assumptions reflected in how \textit{societal impacts} are conceptualised and constructed through ML evaluations.\footnote{Societal `impacts' is the phrasing generally adopted in responsible AI literature, which we follow. `Impact', however, is suggestive of immediacy and collisions. (Mateescu and Elish have made a similar point about `deploy' \cite{mateescu2019ai}.) Social `outcomes' may be preferable. This encourages thinking about long-term and second- and third-order outcomes of introducing AI systems into society.} Doing so reveals that the work of assessing and managing societal impacts of GenAI is best conceptualised through a governance, rather than a prediction, frame.\looseness=-1

Evaluating GenAI's societal impacts requires a model of how these impacts manifest \cite{martin2020ExtendingMachineLearning}. This model, often implicit, enables interpretation of how GenAI systems interact with people and social structures; the model constructs particular social factors as capable of, and deserving of, measurement. We ask: what is the model of societal impacts reflected in existing efforts to evaluate GenAI systems? One avenue to understand this model is to look at taxonomies of societal impacts \cite[\textit{e.g.},][]{solaimanEvaluatingSocialImpact2023a, shelby2023sociotechnical, weidinger2022taxonomy} which provide conceptual infrastructure for societal impact evaluations of GenAI \cite[for an alternate approach see][]{rubambiza2024SeamWorkSimulacra}. We note that some taxonomies refer to societal risks or harms, rather than impacts. Whilst we are wary of conflating these terms \cite{boyd2024RisksVsHarms}, in this provocation, we use ``taxonomies of societal impacts'' as a catchall category for efforts to develop a classification or categorisation system for relating GenAI to social phenomena.

\section{Taxonomies of societal impacts}

Taxonomies of societal impacts exist for a range of GenAI technologies and components, including foundation models \cite{dominguez_hernandez_mapping_2024}, text-to-image models \cite{bird_typology_2023}, large language models \cite{weidinger2022taxonomy}, speech generation models \cite{hutiri_not_2024}, AI agents \cite{chan2023harms}, and GenAI or algorithmic systems more generally \cite{solaimanEvaluatingSocialImpact2023a, shelby2023sociotechnical}. These tools direct the attention of GenAI evaluators, and provide structure for GenAI evaluations, in at least three ways. First, taxonomies of societal impacts enable researchers and practitioners to think systematically about the potential consequences of deploying GenAI technologies. Taxonomy development is, therefore, ontological work that has far-reaching consequences for the way researchers and practitioners understand the relationship between GenAI technologies, people, and society \cite{moss2021AssemblingAccountabilityAlgorithmic, bowkerSortingThingsOut2000}. 
Indeed, the development of bespoke taxonomies for different GenAI technologies implies a conceptualisation of societal impacts that centres technology as the primary causal determinate of harms and positions technology developers as the critical actors in impact evaluations (for contrast, imagine a range of taxonomies for different social contexts).\looseness=-1

Second, taxonomies of societal impacts direct attention by navigating the tension between abstraction and contextualisation, which is present throughout AI development \cite{selbstFairnessAbstractionSociotechnical2019}. GenAI components tend to be understood within an abstract space that disregards the social context of their development \cite{martin2020ExtendingMachineLearning}. Social harms, meanwhile, are understood as being deployment, \emph{i.e.}, context, dependent \cite[\textit{e.g.},][]{solaimanEvaluatingSocialImpact2023a}. \citet{shelby2023sociotechnical}, in their taxonomy, attempt to navigate this tension by distinguishing between harms that originate with computational components of AI systems, and harms that originate in their deployment.
Yet, GenAI systems (\textit{e.g.}, ChatGPT) are deployed extremely broadly, cutting across vast swathes of different social contexts. In this context, taxonomies direct attention towards forms of harm that can be tested or detected at the abstract level of the GenAI model, within the broader software evaluation paradigm in which GenAI deployment occurs \cite[\textit{e.g.}, GPT4,][]{openai_gpt-4_2024}.

Third, taxonomies of societal impacts also direct attention towards prediction and trade-offs, centring the site of GenAI development rather than application or deployment contexts. Taxonomies are often framed as predictive tools, enabling practitioners to forecast risks of GenAI deployment \cite[\textit{e.g.},][]{weidinger2022taxonomy, bird_typology_2023}. Yet, in attempting to develop an exhaustive schema of potential impacts, taxonomies invite comparison, and trade-offs, across disparate categories of societal impacts \cite{douglas1983RiskCultureEssay}, 
such as ``trust in media and information'', ``community erasure'', and ``intellectual property and ownership'' \cite{solaimanEvaluatingSocialImpact2023a}. 
Implicit in this organisation is a conceptualisation of societal impacts as modular, independent, and commensurable, with GenAI developers positioned as arbitrators in determining which impacts to address, and how.\looseness=-1

\section{Conceptualising the societal impacts of GenAI}

A conceptualisation of societal impacts that centres GenAI technologies and GenAI developers may be useful, in terms of producing a discourse on social outcomes that is tractable within GenAI, but should be approached with caution. Critical questions to consider include: what factors should be centred when thinking about GenAI's societal impacts? what are the limits of societal impact prediction? how should evaluators balance different forms of societal impacts? To begin responding to these questions, we offer three premises for rethinking the relationship between GenAI, people, and society.\looseness=-1

Societal impacts should be understood as application- and context-specific \cite{suchman1999reconstructing, martin2020ExtendingMachineLearning} and indeterminate \cite{wynne1992UncertaintyEnvironmentalLearning}. Failure to do so produces an understanding of societal impacts that is universalising and self-fulfilling; the work of evaluating societal impacts becomes the work of extending patterns of social relations from one place to many \cite{haraway1988situated}. Societal impacts of GenAI should be thought of at the system level, with the GenAI system situated in a particular social context \cite[cf., regarding model explanations,][]{smart_beyond_2024}. Impacts manifest when a model is integrated into a sociotechnical system, and implemented in a specific social setting \cite{shelby2023sociotechnical}. Context-specificity makes predictions about societal impacts difficult and unverifiable. Consequently, taxonomies of societal impacts are inherently partial, always incomplete.\looseness=-1

Some societal impacts should be understood as incommensurable \cite{asveld_incommensurability_2009}. The scaling of large multilingual models to include many languages, including Indigenous languages, illustrates this phenomenon \cite{PratapScaling2023,ImaniGooghariGlot2023}. Such scaling is motivated by the assumption that language technologies should be accessible to everyone in their first language \cite{BirdDecolonising2020}, which leads to model evaluations focused on identifying and rectifying performance disparities across languages. Yet, how should evaluators reconcile issues of disparate performance with issues of Indigenous data sovereignty, given one strategy to improve GenAI performance is to collect more language data? Navigating these trade-offs is particularly problematic when the objectives of GenAI developers may diverge from those of local communities. Some Australian Aboriginal and Māori communities
prioritise managing cultural knowledge, including language data, to support intergenerational transmission rather than expanding access to language technology \cite{CooperIts2024,Te_Mana_Raraunga2016}. In contexts like these, while taxonomies of societal impacts can help GenAI practitioners identify a broad range of impacts, GenAI practitioners are not well-positioned to balance competing impacts--these are value-laden decisions that require community leadership. Taxonomies can support such leadership by enabling practitioners to identify relevant stakeholders associated with different societal impacts \cite{bird_typology_2023}.

Finally, questions of societal impacts are questions about social power. Taxonomies of societal impacts enable evaluators to decide what to include (and exclude) in their evaluations. This legitimises particular concerns or forms of impact as salient to GenAI. As evaluation practices mature and become standardised, they gain efficacy, in terms of their capacity to enforce the values, simplifications, and assumptions they reflect \cite{lamont2012comparative, scott2020SeeingStateHow}. The dominance of cost-benefit analysis in environmental impact evaluation, for example, supports a capitalist and extractive epistemology, in which the worth of the environment is expressed in monetary terms \cite{winnerWhaleReactorSearch1986}. 
Efforts to standardise societal impact evaluation---worthy as they are---should, therefore, be understood as sociopolitical efforts that can reify or resist particular social orders. Who determines which societal impacts to focus on matters.

\section{For a governance-first approach}

The conceptualisation of societal impacts sketched above suggests redirecting efforts away from evaluations of potential harms and towards a governance-first approach to GenAI oversight. If societal harms are contextually contingent and indeterminate, then anticipatory evaluations may not be as effective at identifying and mitigating impacts as robust governance and monitoring of GenAI deployment led by stakeholders or governments.
Reflecting this, a governance-first approach would demand accountability to, participation of, and deliberation within, stakeholders or communities impacted by GenAI deployments \cite{RakovaAlgorithms2023}--for example, to determine how to balance disparate impacts. While practical recommendations for implementing a governance-first approach are beyond the scope of this provocation, a helpful starting place is to reconsider the role of taxonomies in RAI evaluations. Taxonomies and other evaluation tools can serve as useful inputs to robust governance and accountability processes \cite{moss2021AssemblingAccountabilityAlgorithmic}. However, without first establishing sustainable, representative governance structures (or engaging with those that already exist), these tools risk generalising predictions of harms across diverging contexts, equating incommensurable impacts, and ultimately serving the interests of GenAI researchers and developers rather than affected communities.\looseness=-1

\begin{ack}
We thank Jochen Trumpf, Kathy Reid, and Timothy Neale for timely literature suggestions, and the anonymous reviewers for their thoughtful feedback. Glen Berman and Ned Cooper are supported by Australian Government Research Training Program (RTP) Scholarships.
\end{ack}

\bibliographystyle{plainnat}

\bibliography{references}

\end{document}